\documentclass[onecolumn,article]{IEEEtran}
\usepackage{amsmath,amsfonts,amssymb,amsthm}
\usepackage{graphicx}
\usepackage{lineno}
\usepackage{amssymb}
\usepackage{pgfplots}
\pgfplotsset{compat=1.3}
\usepackage{amsmath}
\usepackage{graphicx}
\usepackage{upgreek}
\usepackage{comment}
\usepackage[position=top]{subfig}
\usepackage{array}
\newcolumntype{P}[1]{>{\centering\arraybackslash}p{#1}}
\newcolumntype{M}[1]{>{\centering\arraybackslash}m{#1}}
\usepackage{tablefootnote}
\usepackage{siunitx}
\usepackage{soul}

\pgfplotsset{
        colormap={parula}{
            rgb255=(53,42,135)
            rgb255=(15,92,221)
            rgb255=(18,125,216)
            rgb255=(7,156,207)
            rgb255=(21,177,180)
            rgb255=(89,189,140)
            rgb255=(165,190,107)
            rgb255=(225,185,82)
            rgb255=(252,206,46)
            rgb255=(249,251,14)
        },
    }

\title{Parabolic dielectric reflector for  extreme on-chip spot-size conversion with broad bandwidth}

\author{Laureano Moreno-Pozas, Miguel Barona-Ruiz, Robert Halir, José de-Oliva-Rubio, Jorge Rivas-Fernández, Iñigo Molina-Fernández, J. Gonzalo
Wangüemert-Pérez, Alejandro Ortega-Moñux}

\begin{document}

\maketitle
\begin{abstract}
Spot-size converters are key  for efficient coupling of light between waveguides of different sizes. While adiabatic tapers are well suited for small size differences, they become impractically long for expansion factors around ${\times100}$ which are often required when coupling integrated waveguides and  free-space beams.
Evanescent couplers and bragg deflectors can be used in this scenario, but their operation is inherently limited in bandwidth. Here we propose a solution based on a parabolic dielectric interface that couples light from a {${0.5\,\upmu\rm{m}}$}-wide waveguide to a {${285\,\upmu\rm m}$}-wide waveguide, i.e. an expansion factor of ${\times 570}$. We experimentally demonstrate  an unprecedented bandwidth of more than ${380\ \rm{nm}}$ with insertion losses below $0.35\ \rm{dB}$. We furthermore provide analytical expressions for the design of such parabolic spot-size-converters  for arbitrary  expansion factors. 
\end{abstract}

Silicon photonics is a burgeoning technology for optical communications due to its compatibility with complementary metal-oxide-semiconductor (CMOS) processes, which enables the integration of photonic circuits and microelectronics \cite{shekhar2024roadmapping, chen2018emergence, rinaldi2023space}. Spot-size converters (SSC) are key components in such photonic systems as they enable efficient coupling between waveguides of different sizes. 
 In particular, SSCs are often used  to feed gratings that couple light to fibers \cite{halir2009waveguide},  or project it into free-space \cite{kim2018photonic, yulaev2019metasurface, yulaev2022surface, ding2020metasurface}.
 The latter {requires}  spot-size expansions to hundreds of micrometers to achieve  beams with Rayleigh ranges of the order of centimeters \cite{yulaev2022surface}. This is achieved by first using a SSC to expand light on-chip in the lateral direction from a single-mode waveguide to the desired width (typically exceeding $100\ \upmu\rm m$). The resulting expanded mode then feeds a weak grating that gradually couples the light into free-space, thereby providing the desired longitudinal expansion \cite{kim2018photonic}. In this scenario traditional adiabatic tapering cannot provide the required spot size conversion within reasonable footprints and therefore other solutions are preferred in the literature. 

Silicon-nitride (SiN) platforms have leveraged evanescent couplers to feed very large gratings \cite{kim2018photonic, yulaev2019metasurface, yulaev2022surface, ropp2023integrating}. However, in these SSCs the propagating angle of the expanded modes varies with wavelength. This is due to variations in the effective index of the mode traveling through evanescently coupled waveguide and leads to oblique incidence on the grating which causes spurious changes in its radiation angle \cite{kim2018photonic}. In the Silicon-on-insulator (SOI) {platform},  SSC designs based on Bragg deflectors \cite{hadij2019distributed, hadij2021high, ginel2022chip}  have been presented. However, due to their resonant nature these SSCs are also strongly wavelength dependent. Integrated lenses  could potentially overcome these bandwidth limitations, but so far designs have been limited to MFD below  $40\ \upmu\rm m$ \cite{luque2019ultracompact, zhang2020ultra}.

%the input photonic wire, which produces an oblique incidence at the grating input that causes a deviation in its azimuth angle of radiation . 
%In other scenarios where light is coupled into the fundamental mode of a wide waveguide, this effect lead to significant losses since (i) the slab mode may impinge on the waveguide at a point with a high offset from its center, and (ii) even small angular deviations accumulate large phase differences between the very wide fundamental mode of the output waveguide and the propagating slab mode.

%Silicon-on-insulator (SOI) platform solutions have also leveraged metamaterial-based bragg deflectors \cite{hadij2019distributed, hadij2021high, ginel2022chip} and integrated lenses \cite{luque2019ultracompact, zhang2020ultra}, yet they do not generate output modes with MFD surpassing $40\ \upmu\rm m$ \cite{hadij2019distributed, hadij2021high, ginel2022chip, luque2019ultracompact, zhang2020ultra}. Moreover, adapting those metamaterial-based structures to produce larger beams will undoubtedly incur in changing their geometry.

\begin{figure}[tb]
\centering
\subfloat[]{
\centering
\includegraphics[trim={3cm 3cm 0cm 2cm},clip,scale=0.3]{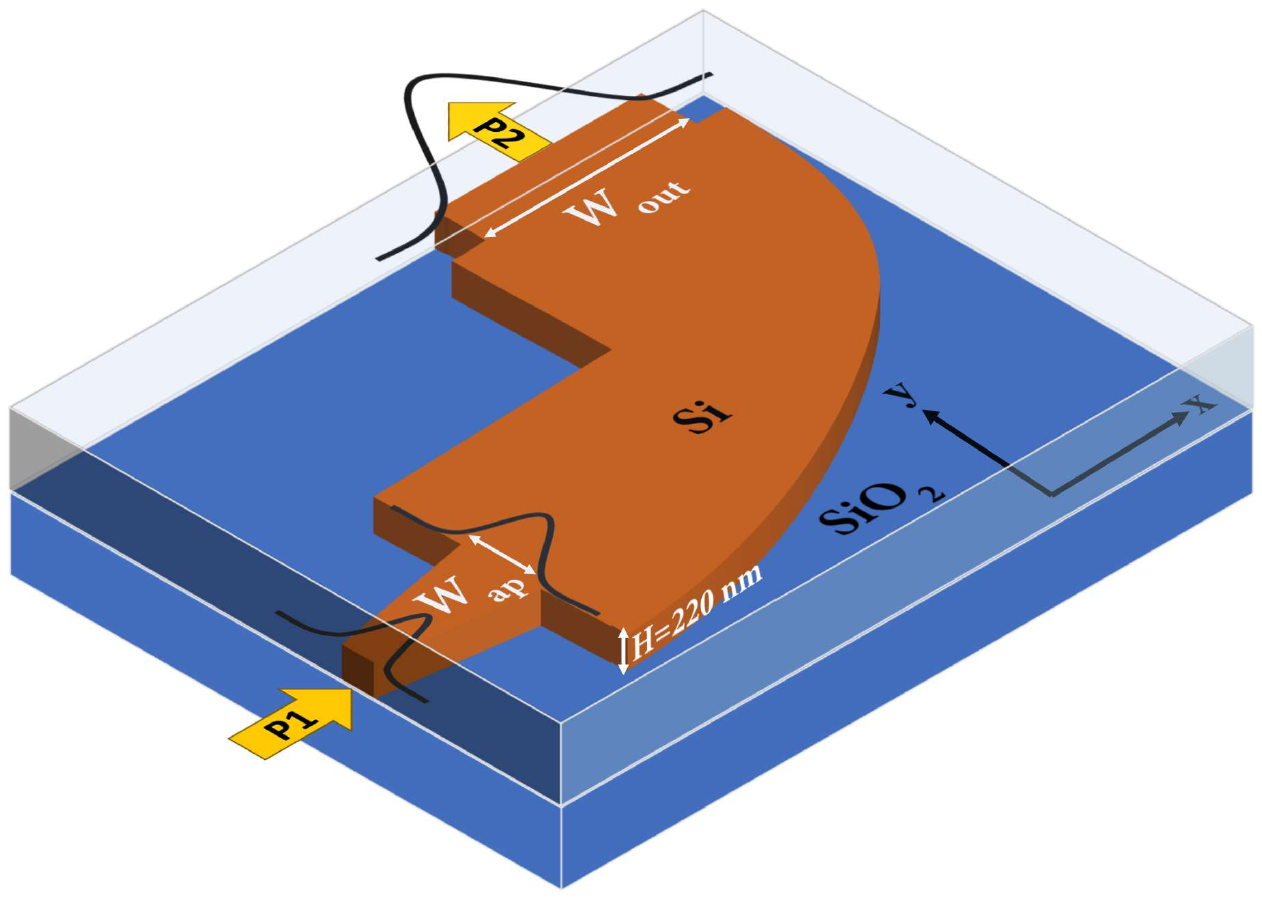}
}

\subfloat[]{
    \centering
\includegraphics[trim={0cm 0cm 9.5cm 0cm},clip,scale=0.32]{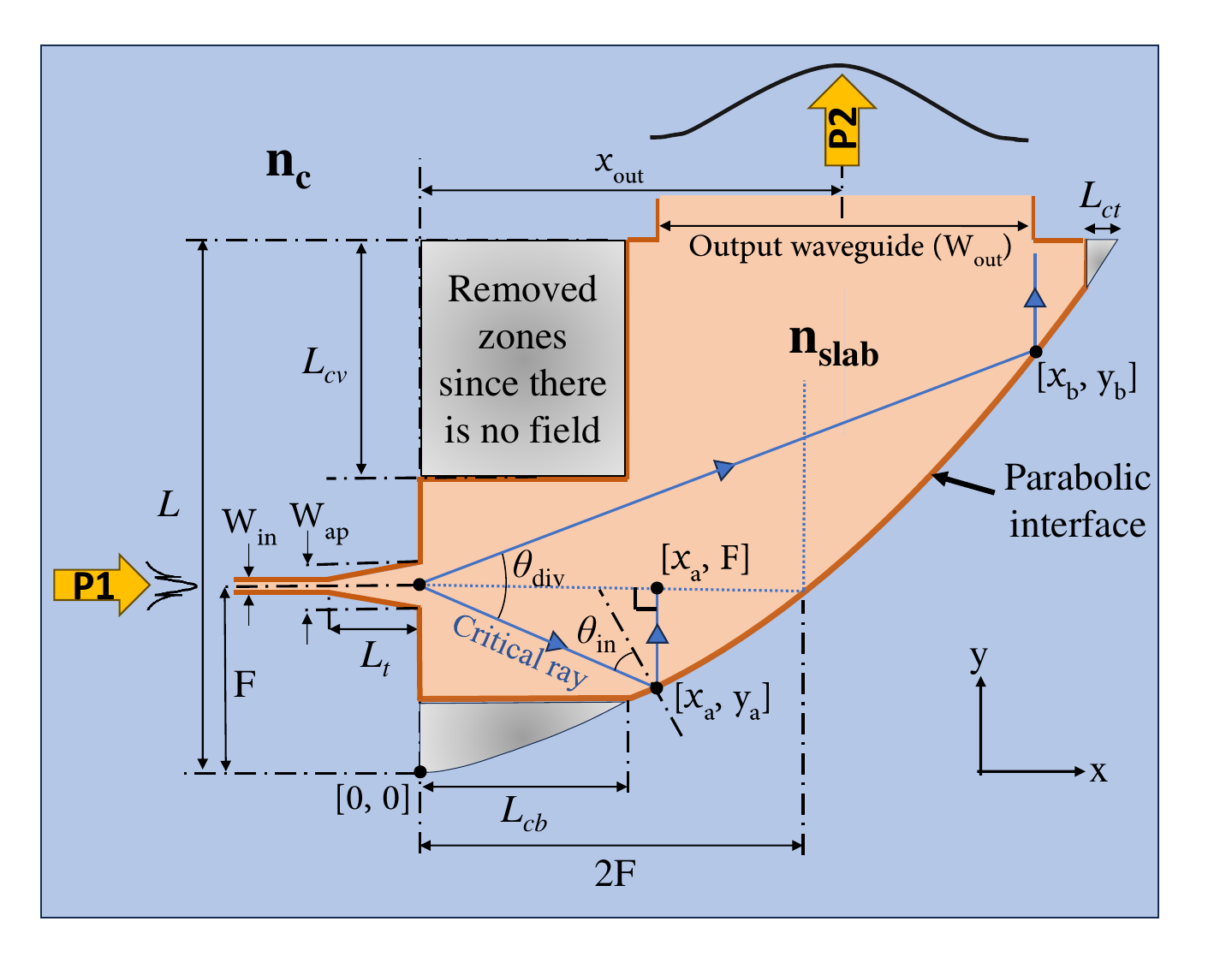}
\label{fig:geometryb}
}
\caption{Device geometry in (a) 3D and (b) 2D. Light diffracts from the aperture of width $W_{\rm{ap}}$ and then impinges the parabolic interface to create a light beam that travels towards the output waveguide.}
\label{fig:geometry}
\end{figure}

An interesting approach based on parabolic dielectric reflectors in SOI was presented in \cite{xu2023compact}. Based  on total internal reflection, an expansion from a single mode waveguide to a MFD of $7\ \upmu\rm m$ was demonstrated, with losses below $0.25\ \rm{dB}$ in a bandwidth of $100\ \rm{n m}$. Here, we show, for the first time, how this approach can be exploited for extreme SSCs, as illustrated in \mbox{Fig. \ref{fig:geometry}}. We derive analytical expressions to i) minimize power loss due to imperfect total internal reflection , and (ii) minimize the area of the device. Using this approach we demonstrate coupling from a the fundamental TE mode of a $0.5\ \rm{\upmu m}$ wide waveguide to the fundamental mode of a $285\ \rm{\upmu m}$ wide waveguide, with measured losses below $0.35\ \rm{dB}$ across a bandwidth exceeding $380\ \rm{nm}$, limited by the measurement setup. In terms of the bandwidth $\times$ expansion figure of merit, this is an $98$-fold improvement with respect to the state-of-the-art. {We believe that this broadband behavior paves the way for the development of different applications, including free-space projection, grating couplers, far-field trapping sensors and wireless chip interconnects, in which one of the limiting performance factors is the reduced bandwidth of conventional beam expanders.}

The geometry of the SSC is depicted in Fig. \ref{fig:geometry}(a) and (b), expanding the fundamental TE mode from a $W_\mathrm{in}=0.5\ \upmu\rm{m}$ wide photonic wire with a thickness of $220\ \rm{nm}$ to an output waveguide with a width of $W_{\rm{out}}$. Both the top and bottom claddings are silicon dioxide, with a refractive index $n_c=1.44$ at $\lambda_0=1.55\,\upmu\mathrm{m}$. Its operation is analogous to an offset parabolic antenna, with the dielectric interface acting as reflector with a focal length $F$. The parabolic interface follows the equation \cite{xu2023compact}
\begin{equation}
    y=\frac{x^2}{4F}.
    \label{eq:parabolic}
\end{equation}
The input waveguide at port $P1$ is adiabatically tapered over a length $L_t$ to a width $W_{\rm{ap}}$, as shown in Fig.~\ref{fig:geometryb}. Light then travels through a free propagation region, experiencing diffraction with a divergence angle $\theta_{\rm{div}}$, until it reflects at the parabolic interface and eventually reaches the output port with width $W_{\rm{out}}$.

The aperture size $W_{\rm{ap}}$ governs the divergence angle as
\begin{equation}
\label{eq:theta_div}
    \theta_{\rm{div}}= \frac{{2.35}\lambda_0}{W_{\rm{ap}} n_{\rm{slab}}},
\end{equation}
where $\lambda_0$ represents the vacuum wavelength and $n_{\rm {slab}}$ denotes the effective index of the free space region ($n_{\rm {slab}}=2.85$ for TE polarization at $\lambda_0=1.55\ \upmu\rm m$). Our definition of  the divergence angle
corresponds to {$\times 1.3$} its classical value  \cite{saleh2019fundamentals}, thereby
encapsulating more than {$95\%$} of the transmitted power within the resultant light cone,  depicted in Fig. \ref{fig:geometryb} with its extreme rays in blue. This is crucial to ensure low-loss operation. The focal length $F$ is then given by \cite{xu2023compact}
\begin{equation}
    F=\frac{W_{\rm{out}}}{4\tan\left({\theta_{\rm{div}}}/{2}\right)}.
    \label{eq:focal_length}
\end{equation}
 The extreme rays in blue in Fig.~\ref{fig:geometryb} impinge on the parabolic interface at points $[x_a, y_a]$ and $[x_b, y_b]$ with coordinates \cite{xu2023compact}
\begin{equation}
x_{a,b}=2F\left[\mp\tan\left(\frac{\theta_{\rm{div}}}{2}\right)+\sqrt{\tan^2\left(\frac{\theta_{\rm{div}}}{2}\right)+1}\right],
\label{eq:x_a}
\end{equation}
and $y_{a,b}$ given by eq.~(\ref{eq:parabolic}).

{To minimize the device area}, we need to maximize the divergence angle or, equivalently, from eq.~(\ref{eq:theta_div}), minimize $W_{\rm{ap}}$. However, $W_{\rm{ap}}$ should not be too small since we must meet total internal reflection (TIR) at the parabolic interface for all the rays coming from the aperture, i.e. ${\theta_{\rm{in}}\geq\arcsin({n_c}/{n_{\rm{slab}}})}$ \cite{orfanidis}. The most restrictive ray is the one that impinges the parabolic interface at $[x_a, y_a]$ since it does so with the smallest angle $\theta_{\rm{in}}^{\min}$. We refer to it as critical ray in Fig.~\ref{fig:geometryb}. From the rectangular triangle conformed by the points $[x_a,y_a]$, $[0, F]$ and $[x_a, F]$ and depicted in Fig \ref{fig:geometryb}, we arrive at the following expression for $\theta_{\rm{in}}^{\min}$:
\begin{equation}
   \theta_{\rm{in}}^{\min}=\pi/4-\theta_{\rm{div}}/4.
\end{equation}
Since the angle $\theta_{\rm{in}}^{\min}$ must meet the TIR condition  \cite{orfanidis}, 
we find that
\begin{equation}
\theta_{\rm{div}}<\pi-4\arcsin\left(\frac{n_{c}}{n_{\rm{slab}}}\right),
\label{eq:appendix_deriv}
\end{equation}
which requires that $\arcsin\left({n_{c}}/{n_{\rm{slab}}}\right)<\pi/4$ since ${\theta_{\rm{div}} >0}$, i.e. ${{n_{\rm{c}}}/{n_{\rm{slab}}}}<1/\sqrt{2}\approx 0.71$. Substituting the divergence angle $\theta_{\rm{div}}$ from (\ref{eq:theta_div}) into (\ref{eq:appendix_deriv}) we find:
\begin{equation}
\label{eq:ec_bonita}
    W_{\rm{ap}} > \frac{{2.35}\lambda_0}{n_{\rm{slab}}\left(\pi-4\arcsin\left(n_{\rm{c}}/n_{\rm{slab}}\right)\right)}.
\end{equation}
For the standard SOI platform with a silicon thickness of $220\ \rm{nm}$ and silica in the top and bottom-cladding ($n_c=1.444$), we have ${n_{\rm{slab}}=2.85}$ for TE polarization {and ${n_{\rm{slab}}=2.05}$ for TM polarization} at $\lambda_0=1.55\ \upmu\rm m$. {For TE polarization, $n_c/n_{\rm{slab}} = 1.44/2.85\approx 0.51 < 0.71$, so the TIR condition} (\ref{eq:appendix_deriv}) {is fulfilled for a wide range of divergence angles (up to $58$ degrees)}{, corresponding to aperture widths $W_{\rm{ap}} > 1.3\ \upmu\rm m$. For TM polarization, $n_c/n_{\rm{slab}}=1.44/2.05\approx 0.7$, which strongly constrains the divergence angle to be less than} $1$ degree{ and the aperture width wider than 126 um, thus hindering the compactness of the device.}
{To overcome this problem, the $n_c/n_{\rm{slab}}$ ratio must be increased, for instance, by removing the silica cladding from the right side of the parabolic reflector. With this approach, $n_c/n_{\rm{slab}} = 1/2.05\approx 0.5$, leading to dimensions similar to those of TE polarization. In the following, we will focus on TE polarization with silica upper cladding.}

Our design methodology  relies on first choosing the aperture size $W_{\rm{ap}}$ to minimize the area of the device, employing the TIR condition proposed in eq.~(\ref{eq:ec_bonita}) with a small margin of $0.2\ \upmu\rm m$, i.e. $W_{\rm{ap}}=1.5\ \upmu\rm m$. Note that even using a slightly larger aperture $W_{\rm{ap}}=2\ \upmu\rm m$ would already result in $30\,\%$ larger device footprint. { The footprint can be further reduced by $8\%$ when selecting $W_{\rm{ap}}=1.3\ \upmu\rm{m}$, although this comes with a $0.4\ \rm{dB}$ increase in insertion losses (IL) in the proposed design.} In this work, insertion losses are defined as $\mathrm{IL}[\mathrm{dB}]=-10\log_{10}(P_2/P_1)$, with $P_2$ the power in the  fundamental TE mode of the output port , and $P_1$ the power of the fundamental TE mode launched into the input port. 

Subsequently, we set the focal length $F$ using equations~(\ref{eq:theta_div}) and (\ref{eq:focal_length}). With these dimensions, aided by Fig. \ref{fig:geometryb}, the parabolic reflector is defined when selecting $L=1.2 y_b$, with $y_b$ the ordinal coordinate of the upper extreme point in the parabolic interface represented in Fig.~\ref{fig:geometryb}, with coordinates $[x_b, y_b]$ given by eq.~(\ref{eq:parabolic}) and eq.~(\ref{eq:x_a}). The factor $1.2$ leaves a small margin of $20\%$ on top of the light cone. The rest of additional equations for delineating the zones where there is no field are given by
\begin{equation}
    L_{cb}=0.8 x_a,
\end{equation}
\begin{equation}
    L_{ct}=\sqrt{1.2 F y_b}-\sqrt{F y_b},
\end{equation}
\begin{equation}
    L_{cv}=1.2 y_b-2F-\frac{(0.8x_a)^2}{4F},
\end{equation}
where the factors $0.8$ and $1.2$ give margins of $\pm 20\%$ around the light cone. {When removing those margins we degrade insertion losses in $0.1\ \rm{dB}$.}

Notably, this design process is entirely analytical. {In the proposed design, we have optimized the output waveguide center by shifting it $+3\ \upmu\rm m$ from the analytical position ${x_{\rm{out}}=2F=315\ \upmu\rm m}$, which yields a $0.2\ \rm{dB}$ improvement of the insertion losses.} 

%with IL defined as
%\begin{equation}
%    \text{IL}=-20\log_{10}\left\vert\int_{-\infty}^{\infty}E_{\rm{mode}}(x)\left\vert E_{\rm{FPR}}(x)\right\vert e^{-j\phi_{\rm{FPR}}(x)}dx\right\vert,
%    \label{eq:OL}
%\end{equation}

\begin{figure}[!t]
\centering
\subfloat[]{
\centering
\centering
\hspace{70pt}\includegraphics[trim={0cm 0cm 0cm 0cm},clip,scale=1.1]{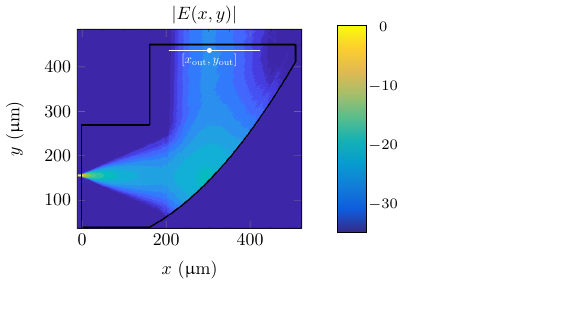}
\label{fig:propagationa}
}
\vspace{-20 pt}

\subfloat[]{
\centering
\hspace{-40pt}\includegraphics[trim={0cm 0cm 0cm 0cm},clip,scale=0.9]{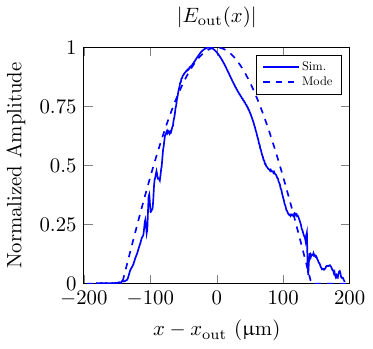}
\label{fig:propagationb}
}
\caption{2.5-FDTD simulation of the  field in the extreme SSC. (a) Propagation of the field and (b) Amplitude of the field at $y=y_{\rm{out}}$. In dashed lines we represent the amplitude of the mode of the output waveguide.}
\label{fig:propagation}
\end{figure}

\begin{figure}[!t]
\centering
\includegraphics[trim={0cm 0cm 0cm 0cm},clip,scale=0.9]{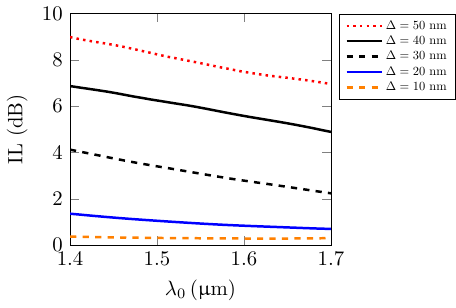}
\caption{Insertion loss for different mesh sizes ${\Delta_x=\Delta_y=\Delta}$ for the extreme SSC as a function of  wavelength, where measurement is given in orange solid line.}
\label{fig:discretization}
\end{figure}

We confirm the analytical design procedure by designing a parabolic SSC with $W_{\rm{out}}=285\ \upmu\rm m$, which corresponds to an expansion factor $W_{\rm{out}}/W_{\rm{in}}$ of $\times 570$, and whose dimensions are given in Table \ref{tab:initial_design}. We perform simulations with Lumerical's 2.5-dimensional finite-difference time-domain (FDTD) method, {since full 3D FDTD simulations of such large devices are unaffordable. As a preliminary check, we have compared the insertion losses calculated with 2.5- and 3-dimensional FDTD simulations for a parabolic reflector with a much smaller footprint ($W_{\rm{out}}/W_{\rm{in}} = 8$). We found that the discrepancy is less than $0.1\ \rm{dB}$ over a wavelength range of $300\ \rm{nm}$, thus confirming the accuracy of our simulation approach}. Upon inspection of the device propagation of field in Fig.~\ref{fig:propagationa}, we confirm that the removed corners exhibit a field drop below $-35\ \rm{dB}$ (dark blue). We also present the magnitude of the field at $y=y_{\rm{out}}$ in Fig.~\ref{fig:propagationb}, which yields $\rm{IL}=0.3\ \rm{dB}$ at $\lambda_0=1.55\ \rm{\upmu m}$. {From Fig.} \ref{fig:propagationb}, {we observe that the electric field is not exactly symmetrical with respect to its maximum, since the photonic parabolic reflector shares the same working principle as offset parabolic antennas commonly used in radio frequencies} \cite{rudge1978offset}. These results are obtained for an FDTD step size of $10\,\mathrm{nm}$ and the time step  given by the Courant formula. This fine mesh is required to avoid errors in the dispersion relation that affect the phase of the lightwave \cite{taflove2005computational}. Indeed, the convergence analysis shown in Fig. \ref{fig:discretization} reveals that larger mesh-sizes fail to accurately describe the ultra-low losses of the device. Decreasing the step size below $10\,\mathrm{nm}$ might be desirable but is limited by practical tradeoffs of memory requirements and simulation time. 

\begin{table}[t]
% increase table row spacing, adjust to taste
\renewcommand{\arraystretch}{1.5}
\setlength{\tabcolsep}{4pt}
\centering
\caption{\bf Design values for the extreme parabolic SSC$^\textit{a}$}
\begin{tabular}
{|c|c|c|c|c|c|c|c|c|c|}
\hline 
$W_{\rm{in}}$ & $L_t $ & $W_{\rm{{ap}}}$ & $W_{\rm{out}}$ & F & L & $x_{\rm{out}}$ & $L_{cb}$ & $L_{ct}$ & $L_{cv}$\\
\hline
\hline
$0.5$ & 10 & $1.5$ & $285$ & 157.5 & 455 & 318 & 203 & 23 & 74.6\\
\hline
\end{tabular}

$^\textit{a}$All values are in micrometers.
\label{tab:initial_design}
\end{table}
%In Fig. \ref{fig:AreavsW1}, we illustrate the expansor area as a function of $W_{\rm{out}}$ for two aperture sizes of $W_{\rm{ap}}$ above the previous TIR condition. It is evident that the device area increases rapidly with higher expansion factors. Additionally, across all expansion factors, $W_{\rm{ap}}=2\ \upmu\rm m$ (the one employed in \cite{xu2023compact}) results in larger expansors compared to $W_{\rm{ap}}=1.5\ \upmu\rm m$. Same situation occurs for $W_{\rm{ap}}=1.5\ \upmu\rm m$ and $W_{\rm{ap}}=1.3\ \upmu\rm m$. Intermediate cases within the range of $W_{\rm{ap}} \in [1.3, 2]\ \upmu\rm m$ behave identically, although they are omitted from the figure for clarity. Therefore, in terms of the area utilization, it is advisable to choose $W_{\rm{ap}}$ as short as possible with $W_{\rm{ap}} > 1.3\ \upmu\rm m$.

%The total silicon area occupied by the parabolic expansor is mainly determined by the desired width of the output port $W_{\rm{out}}$ and the aperture size $W_{\rm{ap}}$ as illustrated in Fig. \ref{fig:AreavsW1}. In the next sections, we will analyze the origin

\begin{figure}[!t]
\centering
\pgfplotsset{every axis/.append style={
xtick={1.2,1.3,..., 1.7},
xmin={1.2},
xmax={1.7},
ymin={0},
ymax={0.4},
}}
\subfloat[]{
    \centering
\includegraphics[trim={0cm 0cm 0cm 0cm},clip,scale=0.35]{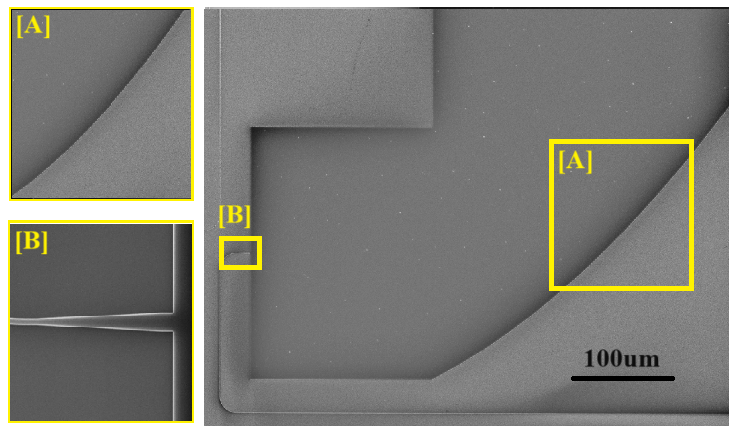}
\label{fig:SEM}
}

\subfloat[]{
\centering
\includegraphics[trim={0cm 0cm 0cm 0cm},clip,scale=0.23]{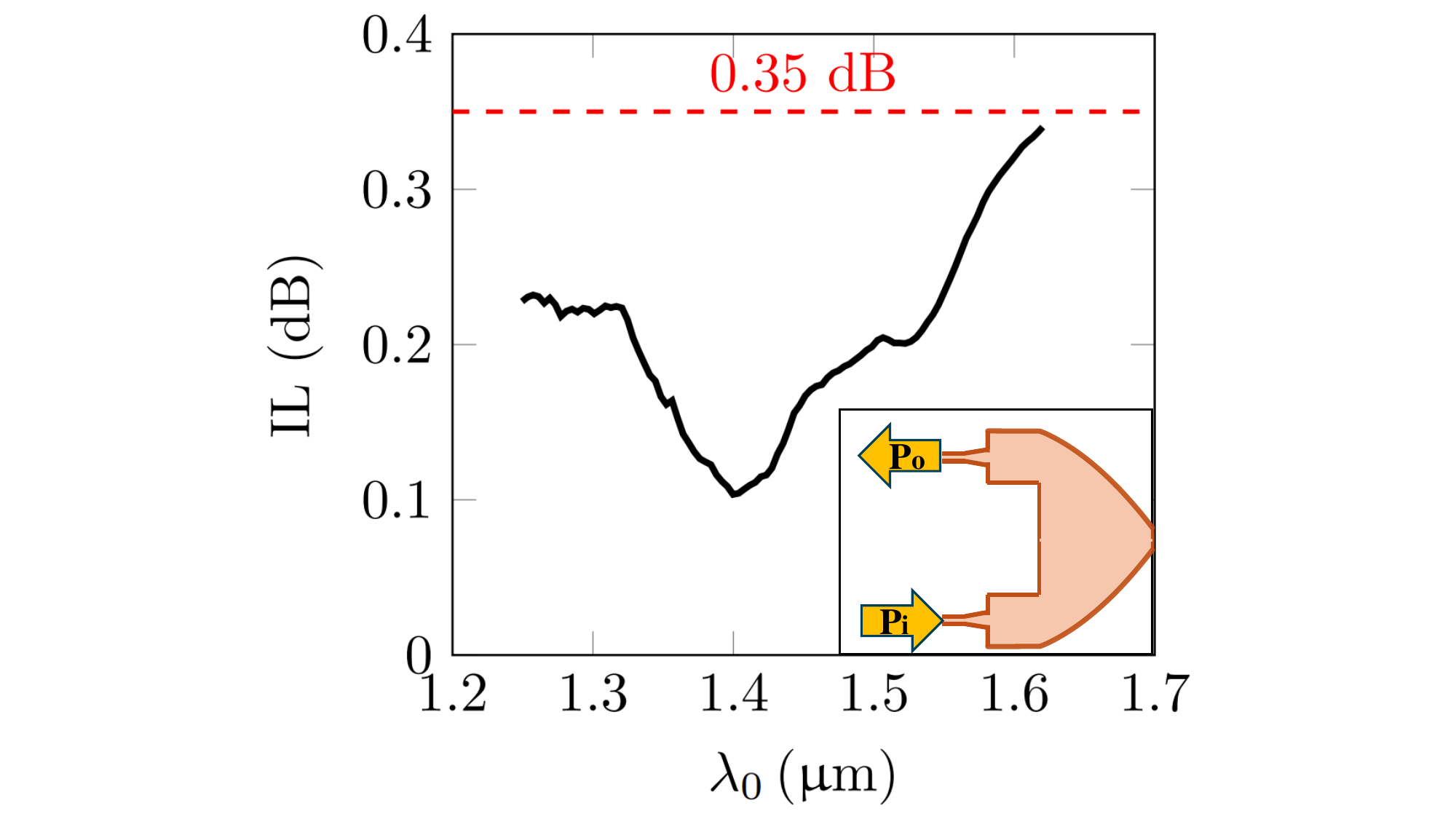}
\label{fig:BW}
}
\caption{Fabricated extreme parabolic SSC. (a) SEM of the device with details and (b) Measured insertion losses of a single SSC. {Insertion losses are measured with the transmission spectrum of the back-to-back configuration shown in the inset.}}
\end{figure}

The extreme SSC with the dimensions specified in Table \ref{tab:initial_design} was fabricated with Applied Nanotools' single etch process, for $220\,\mathrm{nm}$ SOI, with a $2\,\upmu\mathrm{m}$ thick buried oxide layer \cite{ApNanotools}. A $100\,\mathrm{keV}$ electron beam was employed to pattern the resist, and the structures were transferred into the silicon layer through reactive ion etching. A $2.2\,\upmu\mathrm{m}$-thick oxide layer was then deposited on top using chemical vapor deposition.
SEM micrographs of the device are presented in Fig. \ref{fig:SEM} before deposition of the upper cladding. The chip incorporates several flavors of the device with varying positions of the input waveguide (varying $F$ in Fig. \ref{fig:geometryb}) in $\pm 100\rm{nm}$ and $\pm 200\rm{nm}$, to compensate possible fabricating errors in the parabolic interface.

For testing, pairs of identical SSCs were connected back-to-back as shown schematically in the inset of Fig. \ref{fig:BW}, so that insertion losses can be determined with a simple fiber-to-fiber measurement. To cover the $1.26\ \upmu\rm m$ to $1.64\ \upmu\rm m$ wavelength range, light from three tunable lasers (Agilent 81600B series, options 072 and 160, and Santec TSL-770-P-480640-P-F-AP-00-1) is coupled into the chip from a polarization-maintaining fiber though a subwavelength edge coupler \cite{cheben2015broadband}. The fiber is positioned on a rotational stage and combined with a Glan–Thompson polarizer at the output, enabling us to control the polarization coupled into the chip. At the chip's output, light is focused onto a photodetector (Newport 818-IR) using a microscope objective lens. The transmission spectrum is recorded by varying the tunable laser's wavelength while monitoring the output power. 

Measurements of the insertion losses of a single SSC are given in Fig. \ref{fig:BW}. {They correspond to the nominal flavor} ($0\ \rm{nm}$ {offset from the theoretical value of eq.} (\ref{eq:focal_length})). {Devices with input waveguide position offsets of $\pm 100\ \rm{nm}$ and $\pm 200\ \rm{nm}$ exhibited insertion loss degradation at $\lambda_0=1.55\ \upmu\rm{m}$ of up to $0.4\ \rm{dB}$ and $1.2\ \rm{dB}$, respectively. Insertion losses} are obtained by measuring the transmission spectrum of two identical SSCs in the back-to-back configuration shown in the inset of Fig. \ref{fig:BW}, normalizing to a reference waveguide to cancel out fiber-to-chip coupling losses, and dividing the resulting spectrum, expressed in dB, by two. Our measurements demonstrate that $\rm{IL} < 0.35\ \rm{dB}$ from $1.26$ to $1.64\ \upmu\rm{m}$. Our device exhibits an increase of $\times 26$ in expansion and $\times 3.8$ in bandwidth compared to previous parabolic reflectors \cite{xu2023compact}. A detailed comparison of our design with other existing wide-band devices in the literature is provided in \mbox{Table \ref{table_1}}, where the figure of merit (FoM) is given by the product of the $1$-$\rm{dB}$ bandwidth in $\upmu\rm m$ and the magnification. This comparison highlights that our extreme SSC is state-of-the-art and presents an excellent solution for creating SSCs with very large expansion ratios.

\begin{table}[t]
% increase table row spacing, adjust to taste
\renewcommand{\arraystretch}{1.5}
\setlength{\tabcolsep}{1.5pt}
\caption{\bf Performance of some state-of-the art SSC in SOI$^\textit{a}$}
\centering
\begin{tabular}
{c|c|c|c|c|c|c}
\hline 
\text{Ref.} & $W_{\rm{in}}$ & $W_{\rm{out}}$   & \text{IL(dB)} & $\rm{BW}_{\rm{1dB}}(\rm{nm})$ & \text{platform} &\text{FoM}\\
\hline
\hline
\cite{hadij2019distributed, hadij2021high, ginel2022chip} & $0.5$ &$57$  & $0.3$ & \text{N/A} &\text{SOI} &\text{N/A}\\
\hline
\cite{luque2019ultracompact} & $0.5$ &$15$  & $0.6$ & $130$ &\text{SOI} & $3.9$\\
\hline
\cite{zhang2020ultra} & $0.5$ &$10$ &  $0.8$ & $220$ &\text{SOI} & $4.4$\\
\hline
 \cite{xu2023compact} & $0.45$ &$10$ &  $0.15$ & $>100$ & \text{SOI} & $>2.2$\\
 \hline
\text{This work} & $0.5$ &$285$ & $0.23$ & $>380$ & \text{SOI} & $>216.6$\\
\hline
\end{tabular}

$^\textit{a}$All dimensions are in microns unless otherwise specified. Insertion losses are given at $\lambda_0=1.55\ \upmu\rm m$. Bandwidth is measured as the wavelength range where IL are below 1 dB. 
\label{table_1}
\end{table}

In conclusion, we have demonstrated that parabolic dielectric interfaces are a good solution when large spot-size conversions are required. Specifically, we have tackled the conversion of the fundamental TE mode from a $0.5\ \rm{\upmu m}$ SOI waveguide to a waveguide with width up to $285\ \upmu\rm m$, or equivalently a mode with a MFD up to $200\ \upmu\rm m$, i.e. SSC with a magnification factor up to $\times 570$. Key to our result is the TIR condition in Eq.~(\ref{eq:ec_bonita}), which provides a straightforward way to achieve ultra-low losses, below $0.35\,\mathrm{dB}$ over an inherently broad bandwidth ($380\ \rm{nm}$). {The proposed design can be entirely analytical, when fixing the output waveguide center $x_{\rm {out}}=2F$, with no significant cost in terms of IL. This may be of interest to design a Process Design Kit (PDK).} This design represents a promising advancement towards efficient spot-size conversion for different integrated photonic applications, irrespective of the working wavelength.

%\begin{equation}
%    \pi/4-\frac{2\lambda_0}{\pi W_{\rm{ap}} n_{\rm{slab}}}>\arcsin\left(\frac{n_{c}}{n_{\rm{slab}}}\right).
%\end{equation}

\section{Funding} 
We acknowledge funding through project TED2021-130400B-I00/AEI/10.13039/501100011033/ Unión Europea NextGenerationEU/PRTR and project PDC2023-145833-I00/ Ministerio de Ciencia, Innovación y Universidades.

% Bibliography
\bibliographystyle{IEEEtran}
\bibliography{sample}

% Full bibliography added automatically for Optics Letters submissions; the following line will simply be ignored if submitting to other journals.
% Note that this extra page will not count against page length

\end{document}